\documentclass[10pt]{article}
\usepackage{ametsoc2col, verbatim}
\usepackage[hyperindex,breaklinks]{hyperref}

\newcommand{\myabstract}{Several localized versions of the ensemble Kalman filter have been proposed. Although tests applying such schemes have proven them to be extremely promising, a full basic understanding of the rationale and limitations of localization is currently lacking. It is one of the goals of this paper to contribute toward addressing this issue. The second goal is to elucidate the role played by chaotic wave dynamics in the propagation of information and the resulting impact on forecasts. To accomplish these goals, the principal tool used here will be analysis and interpretation of numerical experiments on a toy atmospheric model introduced by Lorenz in 2005. Propagation of the wave packets of this model is shown. It is found that, when an ensemble Kalman filter scheme is employed, the spatial correlation function obtained at each forecast cycle by averaging over the background ensemble members is short ranged, and this is in strong contrast to the much longer range correlation function obtained by averaging over states from free evolution of the model. Propagation of the effects of observations made in one region on forecasts in other regions is studied. The error covariance matrices from the analyses with localization and without localization are compared. From this study, major characteristics of the localization process and information propagation are extracted and summarized.}
\begin{document}

\title{\textbf{\large{On the propagation of information and the use of localization\\in ensemble Kalman filtering}}}
%
% Author names, with corresponding author information. 
% [Update and move the \thanks{...} block as appropriate.]
%
\author{
	\textsc{Young-noh Yoon}
	\thanks{\textit{Corresponding author address:} Young-noh Yoon, Box 234, 082 Regents Dr, Department of Physics, 				University of Maryland, College Park, MD 20742
	\newline{E-mail: mystyle@umd.edu}}\quad
	\textsc{and Edward Ott}\\
	\textit{\footnotesize{Department of Physics, University of Maryland, College Park, Maryland, USA}}
	\and 
	\centerline{\textsc{Istvan Szunyogh}}\\% Add additional authors, different insitution
	\centerline{\textit{\footnotesize{Department of Atmospheric Sciences, Texas A\&M University, College Station, Texas, USA}}}
}
%
% Formatting done here...Authors should skip over this.  See above for abstract.
\ifthenelse{\boolean{dc}}
{
\twocolumn[
\begin{@twocolumnfalse}
\amstitle

% Start Abstract (Enter your Abstract above.  Do not enter any text here)
\begin{center}
\begin{minipage}{13.0cm}
\begin{abstract}
	\myabstract
	\newline
	\begin{center}
		\rule{38mm}{0.2mm}
	\end{center}
\end{abstract}
\end{minipage}
\end{center}
\end{@twocolumnfalse}
]
}
{
\amstitle
\begin{abstract}
\myabstract
\end{abstract}
}
%%%%%%%%%%%%%%%%%%%%%%%%%%%%%%%%%%%%%%%%%%%%%%%%%%%%%%%%%%%%%%%%%%%%%
% MAIN BODY OF PAPER
%%%%%%%%%%%%%%%%%%%%%%%%%%%%%%%%%%%%%%%%%%%%%%%%%%%%%%%%%%%%%%%%%%%%%
\newpage

\let\oldthefootnote\thefootnote
\renewcommand{\thefootnote}{\fnsymbol{footnote}}
\footnotetext[1]{Corresponding author e-mail: \url{mystyle@umd.edu}}
\let\thefootnote\oldthefootnote

\section{Introduction}

% ensemble Kalman filter
In an ensemble Kalman filter (EnKF), at any given analysis time $t=nT$, the estimated state of the atmosphere and the corresponding error covariance reflecting uncertainty in the state estimation are represented by a finite ensemble of global system states \citep[e.g.,][]{Evensen-94, Burgers-et-al-98, Houtekamer-and-Mitchell-98}. Each ensemble member is then integrated forward in time using a physical model of the atmosphere, thus creating an ensemble of forecasts at the next analysis time, $t=(n+1)T$. By suitable incorporation (assimilation) of new measurements, the data assimilation process creates a new ensemble of system states that reflects the most probable atmospheric state and its uncertainty based on the available combined knowledge contained in the forecast ensemble at time $(n+1)T$ and the new measurement data. The process then repeats at the $t=(n+2)T$ cycle, and so on.

% local ensemble Kalman filter
Several localized versions of EnKF have been proposed \citep[e.g.,][]{Houtekamer-and-Mitchell-01, Hamill-et-al-01, Anderson-01, Anderson-07, Ott-et-al-04, Hunt-et-al-07}. In this paper, we concentrate on one localization scheme called the local ensemble Kalman filter \citep[LEKF;][]{Ott-et-al-04} and a subsequent computationally more efficient version, the local ensemble transform Kalman filter \citep[LETKF;][]{Hunt-et-al-07}. We expect that similar results would be obtained using any other of the available EnKF localization schemes. In LEKF and LETKF, local regions surrounding each gridpoint location are defined, the analysis is performed in each local region, and the results are combined to form a global analysis ensemble. The motivation for using localization is that it is expected to work well with drastically fewer ensemble members than would be the case if localization were not employed. In fact, without localization, the computational requirements for a meaningful EnKF analysis are so vast that they make the approach completely infeasible in typical weather forecasting settings. On the other hand, the use of localization has been shown to be successful and eminently feasible for both toy models and real atmospheric models \citep[e.g.,][]{Houtekamer-and-Mitchell-05, Houtekamer-et-al-05, Whitaker-et-al-08, Szunyogh-et-al-08}.

% the first goal
One goal of this paper is to study the basic properties and rationale for localization. To do this, we will employ numerical experiments on a simple model system of equations introduced by Lorenz \citep{Lorenz-05}. The model is simple and small enough that (unlike a real operational model) we can employ an EnKF without localization. We will compare the results obtained with the localized LETKF scheme with those obtained without localization [henceforth referred to as ensemble transform Kalman filter (ETKF); ETKF was introduced in \cite{Bishop-et-al-01}]. By doing this, we will be able to study some aspects relevant to the issues of localization, such as what determines a good size of the localization region, what dynamics justifies (or not) localization, and so on.

% the second goal
The second goal of this paper is to study the propagation of information in a spatiotemporally chaotic system with wave-like dynamics that are similar to that of weather. In particular, we use the traditional approach of discussing the propagation of information based on the study of phase and group velocities of the waves in the model \citep[e.g.,][]{Charney-49, Persson-00, Szunyogh-et-al-02, Zimin-et-al-03}. An alternative approach that has recently been gaining increasing attention is based on using tools of probability theory---most important, measures based on relative entropy of information \citep[e.g.,][]{Kleeman-07}. The relation between localization and the advection of information from the observation location by the model dynamics has been studied using the hierarchical ensemble filter by \cite{Anderson-07}.

% organization of the paper
The organization of this paper is as follows: In section \ref{Model}, we introduce the model system \citep{Lorenz-05} that we employ, and we investigate some of its salient properties. In section \ref{Correlation structure of the model}, we investigate correlations of the model. In section \ref{Forecasting and localization}, we explain the forecast and localization procedures used in the ETKF and LETKF schemes. In section \ref{Spatial correlations of the background ensemble}, we study correlations of the background ensemble of an ETKF as a function of spatial distance. In section \ref{ETKF and LETKF covariance matrices}, we compare covariance matrices obtained using an ETKF and an LETKF. In section \ref{Demonstration of the localized influence of observations}, we show how observations taken in a small region at a certain time affect the forecast at other spatial points in the future. Section \ref{Conclusion} provides further discussions and summarizes our main conclusions.

\section{Model}\label{Model}

% Introduction to Lorenz models
\cite{Lorenz-05} discusses three one-dimensional toy models that incorporate many features shown in real atmospheric dynamics and in global numerical weather prediction models. The first model (Lorenz model 1) was originally introduced in \cite{Lorenz-96} and \cite{Lorenz-and-Emanuel-98}. This model has become the standard model of choice for the initial testing of EnKF schemes. The popularity of the model is in part due to the similarity between the propagation of uncertainties (forecast errors) in Lorenz model 1 and global models in the midlatitude storm-track regions. In particular, the errors are propagated by dispersive waves whose behavior is similar to that of synoptic-scale Rossby waves, and the magnitude of the errors has a doubling time of about 1.5 days (where the dimensionless model time has been converted to dimensional time by assuming that the characteristic dissipation time scale in the real atmosphere is 5 days; see \citealt{Lorenz-96}). Lorenz model 2 adds the feature of a smooth spatial variation of the model variables that resembles the smooth variation of the geopotential height (streamfunction) at the synoptic and large scales in the atmosphere. Lorenz model 3, the most refined and ``realistic'' of the three models in \cite{Lorenz-05}, adds a rapidly varying small-amplitude component to the smooth large-scale flow, mimicking the effects of small-scale atmospheric processes. In our following simulations, we use Lorenz model 3.

% equations of the model
The equation of evolution of the scalar state variable $Z_n$ at position $n$ is the following:
\begin{align}
dZ_n/dt &= [X,X]_{K,n} + b^2 [Y,Y]_{1,n} + c [Y,X]_{1,n}\nonumber\\
&\hspace{11pt} - X_n -b Y_n + F,\label{Lorenz eq}
\end{align}
where $n$ is a spatial variable ($n=1,2,\ldots,N$), periodic boundary conditions are employed ($Z_{N+1}=Z_1$), and the $X$ and $Y$ vectors are defined as
\begin{align}
X_n &= \sum_{i=-I}^{I}{}'(\alpha - \beta |i|)Z_{n+i}, \\
Y_n &= Z_n - X_n, \\
\textrm{with } \alpha &= (3I^2 + 3)/(2I^3+4I), \notag \\
\beta &= (2I^2 + 1)/(I^4+2I^2). \notag
\end{align}
The prime notation on $\Sigma'$ signifies that the first and the last terms in the summation are divided by 2. The bracket of any two vectors $X$ and $Y$ is defined as
\begin{align}
[X,Y]_{K,n} &= \sum_{j=-J}^{J}{}' \sum_{i=-J}^{J}{}' ( -X_{n-2K-i} Y_{n-K-j}\nonumber\\
&\quad + X_{n-K+j-i} Y_{n+K+j} ) / K^2\label{bracket}
\end{align}
when $K$ is even, and $\Sigma'$ is replaced by $\Sigma$ when $K$ is odd; $J=K/2$ when $K$ is even, and $J=(K-1)/2$ when $K$ is odd. The parameter values used throughout this paper are $N = 960, K = 32, b = 10, c = 2.5, F = 15$, and $I = 12$. We solve Eqs.~(\ref{Lorenz eq})--(\ref{bracket}) using a Runge-Kutta scheme with a time step sufficiently small to resolve the fast time scale of the model. We find that these equations lead to a state profile that shows small-scale activity $Y_n$ superposed on a large-scale smooth component $X_n$, thus crudely mimicking the multiscale dynamics of a real atmospheric system.

\begin{figure}
\noindent\includegraphics[width=21.18pc]{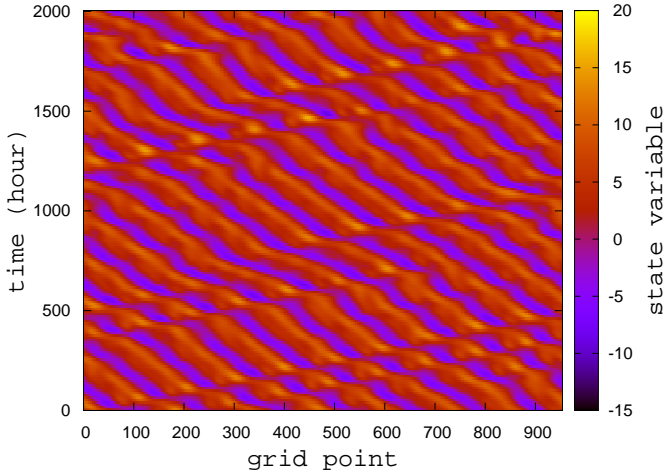}
\caption{Time evolution of state $Z_n$ of Lorenz model 3. In this figure, 6 h are defined to be 0.05 time units in the model equations considering the damping time of the model.}\label{time evolution}
\end{figure}

% wave propagation of the model
Figure \ref{time evolution} obtained from a numerical solution of Eq.~(\ref{Lorenz eq}) shows how states evolve in time. The horizontal axis is the spatial location, and the vertical axis is the time. We define 6 h to be 0.05 time units in the time evolution equation following Lorenz [Lorenz justifies this correspondence between $t$ in Eq.~(\ref{Lorenz eq}) and pseudohours by considering the damping time of the model]. The colors in Fig.~\ref{time evolution} represent the values of the state variables $Z_n$. Red corresponds to high values, and blue corresponds to low values. The figure shows that waves of approximately 7 spatial periods propagate to the left. This observed structure mimics what is seen at large scale in the atmosphere where Rossby waves play a prominent role.

% wave packet propagation of the model
The structure found in Fig.~\ref{time evolution} also suggests that an insight can be gained by representing the field as a modulated sinusoidal wave. For this purpose, modulating envelopes were extracted from the states at each time using the method introduced in the paper by \cite{Zimin-et-al-03}. The method of \citeauthor{Zimin-et-al-03} for extracting modulating envelopes is explained briefly in appendix A. Based on the observed structure in Fig.~\ref{time evolution}, we computed envelopes using wavenumbers of only 7 and 8 to show smooth flow of the envelope (see appendix A). Figure \ref{envelope evolution} depicts the envelope amplitude in time and space for the same numerical run as in Fig.~\ref{time evolution}. From Fig.~\ref{envelope evolution}, we see that wave packets propagate to the right, in contrast to the leftward propagation of the individual troughs and ridges seen in Fig.~\ref{time evolution}. That is, in some appropriate sense, the wave turbulent state consists of waves that have phase velocities that move to the left (as in Fig.~\ref{time evolution}) and group velocities that move to the right (as in Fig.~\ref{envelope evolution}). This situation is similar to that of Rossby waves in the atmosphere, whose phase velocity is always westward relative to the mean zonal flow and whose group velocity can be either eastward or westward.
\begin{figure}
\noindent\includegraphics[width=21.18pc]{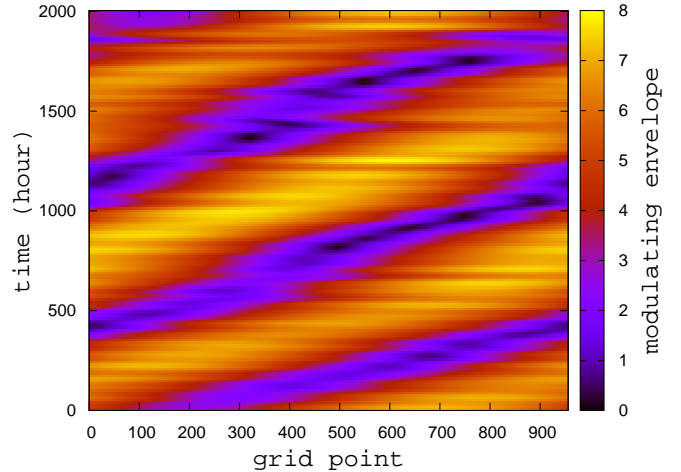}
\caption{Time evolution of the wave packet envelopes extracted by the envelope extraction method introduced in \cite{Zimin-et-al-03}. Wave numbers of only 7 and 8 were used to smooth out the envelopes.}\label{envelope evolution}
\end{figure}

\section{Correlation structure of the model}\label{Correlation structure of the model}

\begin{figure}[!h]
\noindent\includegraphics[width=21.18pc]{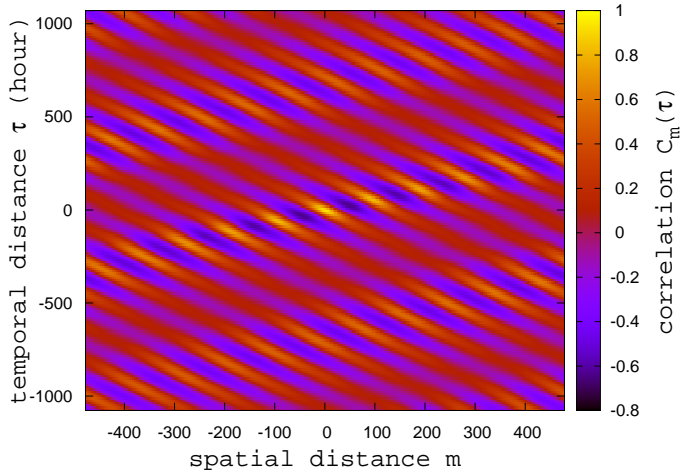}
\caption{Correlations $C_{m}(\tau)$ between state variables at two different points, $(n,t)$ and $(n+m, t+\tau)$ in Fig.~\ref{time evolution}. The values were averaged over all of the space-time locations $(n,t)$.}\label{correlations}
\end{figure}

\begin{figure}[!h]
\noindent\includegraphics[width=21.18pc]{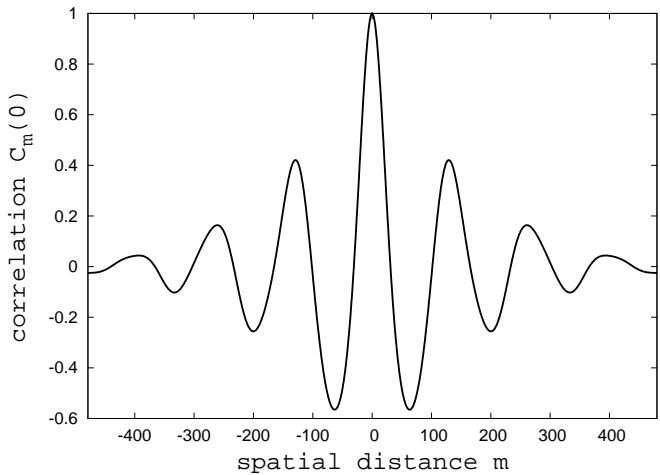}
\caption{Correlations $C_{m}(0)$; i.e., the correlation values of Fig.~\ref{correlations} with time difference $\tau$ set to 0.}\label{correlations-states}
\end{figure}

% correlation of the model
We calculated the correlations $C_{m}(\tau)$ between values at two different locations separated by $m$ grid points in space and by $\tau$ in time as follows:
\begin{equation}
C_m(\tau) = \frac{\langle \, \left[ \, Z_{n}(t)-\langle Z_n(t) \rangle \, \right] \left[ \, Z_{n+m}(t+\tau)-\langle Z_n(t)\rangle \, \right] \, \rangle}{\langle \, \left[ \, Z_{n}(t)-\langle Z_n(t) \rangle \, \right]^2 \, \rangle},\label{eq cor states}
\end{equation}
where angle brackets denote an average over $n$ and $t$. The resulting plots, $C_m(\tau)$ (Fig.~\ref{correlations}) and $C_m(0)$ (Fig.~\ref{correlations-states}), show that wave dynamics plays a central role in the spatiotemporal evolution of the correlation, as demonstrated through three properties: (1) the spatial correlation has the structure of a wave packet with carrier wavenumber 7 and a packet envelope that decreases monotonically with increasing $|m|$ (Fig.~\ref{correlations-states}), (2) the wave structure shifts to the left with a phase velocity of about -0.77 grid points per hour (Fig.~\ref{correlations}), and (3) the packet envelope shifts to the right at a rate that we identify as a group velocity (e.g., see the bright yellow spots along the red troughs in Fig.~\ref{correlations}). Under the assumption of ergodicity, the correlation in Eq.~(\ref{eq cor states}) can be regarded as describing an ensemble drawn from the distribution that defines the climate of the model. In section \ref{Spatial correlations of the background ensemble}, we will consider another type of correlation function that can be regarded as describing an ensemble drawn from a distribution of short-term forecast uncertainties associated with a fairly dense observational network.

\begin{figure}[!h]
\noindent\includegraphics[width=21.18pc]{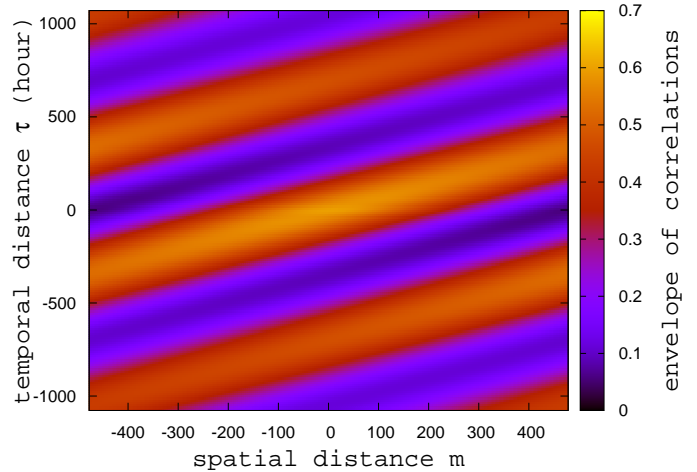}
\caption{Envelopes extracted from correlations in Fig.~\ref{correlations} at each time $\tau$ using wave numbers only from 6 to 8 to smooth out the result.}\label{envelope}
\end{figure}

% envelope of the correlation
To make property (3) more transparent, we extracted the envelopes at each temporal distance $\tau$ from Fig.~\ref{correlations} using the envelope extraction method in \cite{Zimin-et-al-03} with wavenumbers of 6, 7, and 8. We obtained the results depicted in Fig.~\ref{envelope}. It is seen from Fig.~\ref{envelope} that, similar to Fig.~\ref{envelope evolution}, the envelope in $m$-$\tau$ space propagates to the right at a group velocity of about 1.37 grid points per hour.

\section{Forecasting and localization}\label{Forecasting and localization}

% forecast procedure
In our numerical assimilation experiments, we will employ ``perfect model'' simulations of forecast/analysis cycles. That is, we carry out a time evolution run of our model, Eqs.~(\ref{Lorenz eq})--(\ref{bracket}), which we regard as simulating a ``true'' atmospheric state evolution that we wish to analyze and forecast. We then take simulated measurements of this true state $Z_n$ by adding uncorrelated noise of Gaussian distribution to the true state variable $Z_n$ at each grid point and at each analysis time. The mean of the noise is 0 and the standard deviation is 0.3, which is about the size of the small-scale activity $Y$ in the model. The standard deviation of the noise can be also compared with the climatological standard deviation of the state variable $Z_n$, which is 4.67 (a noise standard deviation of 0.3 is used throughout this paper). Using these simulated measurements, we do analyses employing the same model, Eqs.~(\ref{Lorenz eq})--(\ref{bracket}), as we use to generate the true state evolutions [i.e., Eqs.~(\ref{Lorenz eq})--(\ref{bracket})].

% disadvantage of ETKF
In our ETKF implementation, we use the same technique as described in \cite{Hunt-et-al-07}, but without employing localization. To faithfully represent the full system state and its covariance in the absence of localization, the number of ensemble members $k$ that we use in the ETKF must be at least on the order of the number of growing local Lyapunov exponents, which increases as the number of grid points of the model increases in general. Thus, we need many ensemble members when there are many grid points in the model. We integrate each ensemble member during the evolution phase of the forecast cycle, and the required computing power to do this scales like $k$. In carrying out our ETKF procedure, it is also necessary to invert a $k \times k$ matrix, which requires a number of computer operations on the order of $k^3$. Hence, if we can do computations in local regions with much fewer grid points than in the global region, and thus with correspondingly much fewer ensemble members, then the required computational power is greatly reduced. For a more detailed discussion of the computational cost of LETKF, see \cite{Hunt-et-al-07} and \cite{Szunyogh-et-al-08}.

% Motivation and procedure of localization
If the errors in the state estimates at two different grid points that are far from each other are independent, then we might be able to compute analysis ensembles for each location while ignoring the other. This qualitative type of consideration is what motivates localization of the analysis. A brief description of the localization process used in LETKF and implemented on our one-dimensional model, Eq.~(\ref{Lorenz eq}), is the following: First, choose an appropriate spatial size $L$ for the localization. At each grid point $i$, a local patch is assigned as consisting of the $2L+1$ grid points $(i-L, i-L+1, \ldots, i+L-1, i+L)$ for $i=1,2,\ldots,N$. We then compute the analysis ensemble in each local patch using observations in the local patch. Then, we construct the global analysis ensemble that combines all of the local analysis ensembles by taking weighted averages of analysis values from each local patch. \cite{Ott-et-al-04}, \cite{Hunt-et-al-07}, and appendix B explain how to do the steps in this procedure in detail. How accurate is this localization procedure? The supposed main requirement is that observations outside the local patch should not affect the information from the local patch that is used in the evolution of the ensemble to the next analysis cycle. We discuss this issue for Lorenz model 3 in what follows. Note that, because we investigate a univariate model, the issue of balance does not arise. The issue of balance and its interaction with localization has been discussed by \cite{Kepert-09}.

\section{Spatial correlations of the background ensemble}\label{Spatial correlations of the background ensemble}

% Common misunderstanding
As we have pointed out, Fig.~\ref{correlations-states} shows a long-range correlation essentially extending over the entire length of the simulation region. Thus, one might think that we will lose a lot of information that comes through correlations from observations outside the local patch when we localize the analysis scheme. As we shall subsequently argue, this is not the case. In particular, we emphasize that it is the covariance matrix of the background ensemble around the ensemble mean that is used in the analysis, not the covariance matrix of the climatological distribution of true states themselves.

\begin{figure}
\noindent\includegraphics[width=21.18pc]{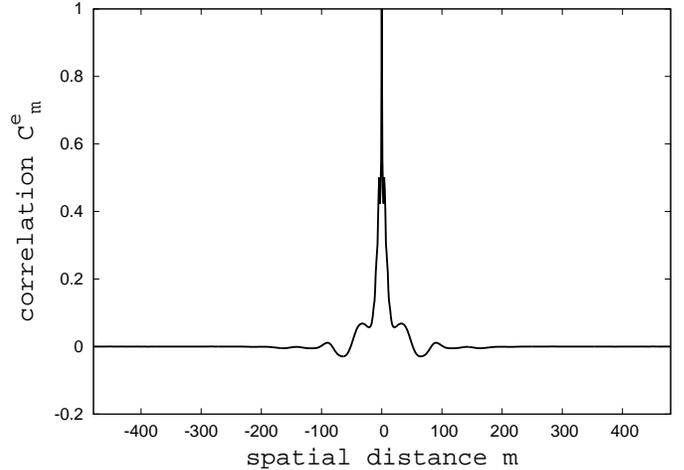}
\caption{Correlation $C_m^e$ of the background ensemble of ETKF with 6-h cycle time vs spatial distance $m$. The values were averaged over 1000 forecast cycles.}\label{correlations-ensemble-06h}
\end{figure}

% Correlations of the background ensemble
To formulate a correlation relevant to what is used in our analyses, we first carry out a perfect-model simulation of a forecasting situation. In particular, we begin by preparing a background ensemble $Z_n^{(k)}(t), k=1,2,\ldots,960$, by running 1000 forecast cycles (we use $k$ and $t$ to denote an ensemble member and cycle time, respectively). During each cycle, we saved the background ensemble $Z_n^{(k)}(t)$ and took simulated observations with noise of uncorrelated Gaussian distribution at each grid point at the analysis time, and then we updated the ensemble with ETKF using a multiplicative covariance inflation factor of 1.13 [see step 5 of appendix B; the factor 1.13 gives the minimum rms error of the resulting state estimate---see \cite{Anderson-and-Anderson-99} for a general explanation of covariance inflation], and we then evolved the true state and the ensemble from $t$ to $t+6$ h (except where otherwise stated, a 6-h cycle time and 1.13 multiplicative covariance inflation factor are used throughout this paper). Using these parameters, we calculated the following ensemble-based spatial correlation in the background:
\begin{align}
C_m^e &= \left< \,\frac{A}{\sqrt{B}\sqrt{C}} \, \right>_{n,t},\label{Cem} \\
\textrm{with } A &= \langle \, [ \, Z_n^{(k)}(t)-\bar{Z}_n(t) \, ] \, [ \, Z_{n+m}^{(k)}(t)-\bar{Z}_{n+m}(t) \, ] \, \rangle_k,\nonumber \\
B &= \langle \, [ \, Z_n^{(k)}(t)-\bar{Z}_n(t) \, ]^2 \, \rangle_k,\nonumber \\
C &= \langle \, [ \, Z_{n+m}^{(k)}(t)-\bar{Z}_{n+m}(t) \, ]^2 \, \rangle_k,\nonumber
\end{align}
where $\langle \cdot \rangle_k$ denotes an average over ensemble members $k$, $\langle \cdot \rangle_{n,t}$ denotes an average over space $n$ and time $t$, and $\bar{Z}_n = \langle \, Z_n^{(k)} \, \rangle_k$. Figure \ref{correlations-ensemble-06h} shows the result. As can be seen, the magnitude of the correlation is less than 3\% of its peak value for $|m|>46$, and the half-width of the strong peak around zero distance is roughly on the order of 10 for 6-h cycle time simulation. Thus, the correlations of deviations of ensemble members from the ensemble mean will be negligible when 2 grid points are separated by more than 46 grid points.

\begin{figure}
\noindent\includegraphics[width=21.18pc]{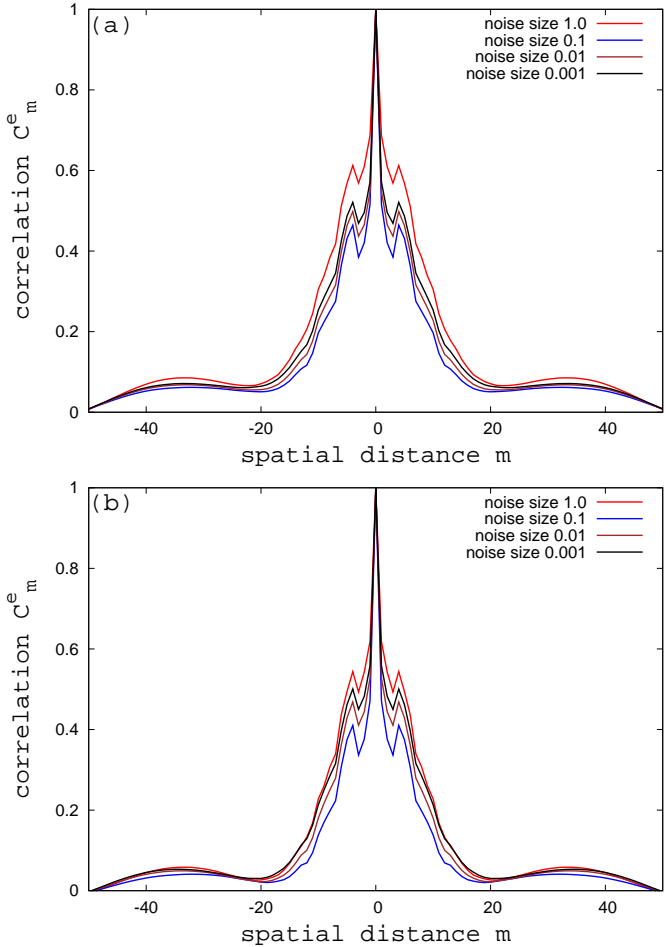}
\caption{Correlations $C_m^e$ (a) of the background ensembles and (b) of the analysis ensembles of ETKF with several different noise sizes. The values were averaged over 1000 forecast cycles. The red, blue, brown, and black lines are for standard deviations of the noise: 1.0, 0.1, 0.01, and 0.001, respectively.}\label{fig-07}
\end{figure}

% Interpretation
To analyze further the result shown in Fig.~\ref{correlations-ensemble-06h}, we note that the background correlation is determined by the structure (correlation) of the analysis error and the model dynamics that spreads the effects of the analysis error in space and time. The shape of the analysis correlation function is very similar to that of the background correlation function and is only weakly dependent on the magnitude of the observation noise (Figs.~\ref{fig-07}a,b). A reasonable assumption is that the spatial expansion of the correlation pattern of the analysis ensemble during the model integration is controlled by the group velocity of the waves that compose the envelope of the analysis error pattern. Although the group velocity associated with these waves is not necessarily identical to the group velocity of the waves observed in the free run of the model, the group velocity observed for the free run can be considered to be an estimate of the maximum group velocity for the given dynamical system. Thus, the comparatively short range of the correlations is natural if one assumes that the expansion of the observed half-width of the correlation pattern of the analysis ensemble during the model integration cannot be larger than the maximum group velocity multiplied by the forecast time used in calculating the values $Z_n^{(k)}(t)$ in Eq.~(\ref{Cem}) and identifies the maximum group velocity (maximum information propagation speed) to be on the order of the speeds found in Figs.~\ref{time evolution}--\ref{envelope} (i.e., $\sim$1 grid point per hour).

\begin{figure}
\noindent\includegraphics[width=21.18pc]{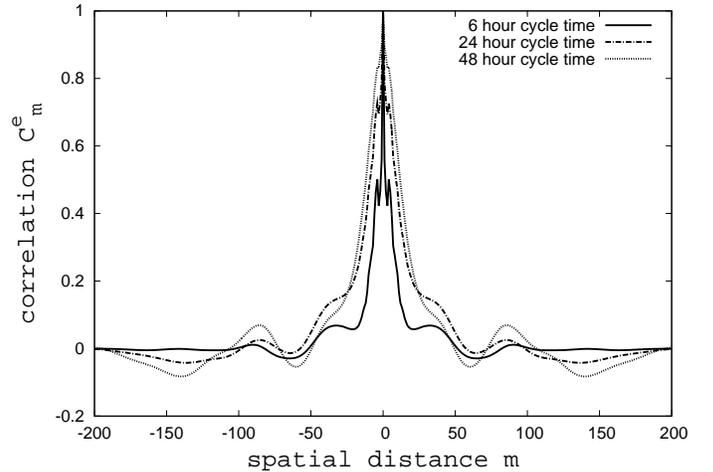}
\caption{Correlation $C_m^e$ of the background ensembles of ETKF with various cycle times vs spatial distance $m$. The values were averaged over 1000 forecast cycles. The solid, dash-dotted, and dotted lines are for 6-, 24-, and 48-h cycle times, respectively.}\label{correlations-ensemble}
\end{figure}

% Various cycle times
In Fig.~\ref{correlations-ensemble}, we show results for the same quantity as plotted in Fig.~\ref{correlations-ensemble-06h}, but for the cases in which the assimilation cycle times are 6 (as in Fig.~\ref{correlations-ensemble-06h}), 24, and 48 h. The observations were taken only at analysis times with corresponding intervals of 6, 24, and 48 h. Consistent with the above interpretation, we observe that the correlation function spreads as the cycle time is increased.

\section{ETKF and LETKF covariance matrices}\label{ETKF and LETKF covariance matrices}

\begin{figure}
\noindent\includegraphics[width=21.18pc]{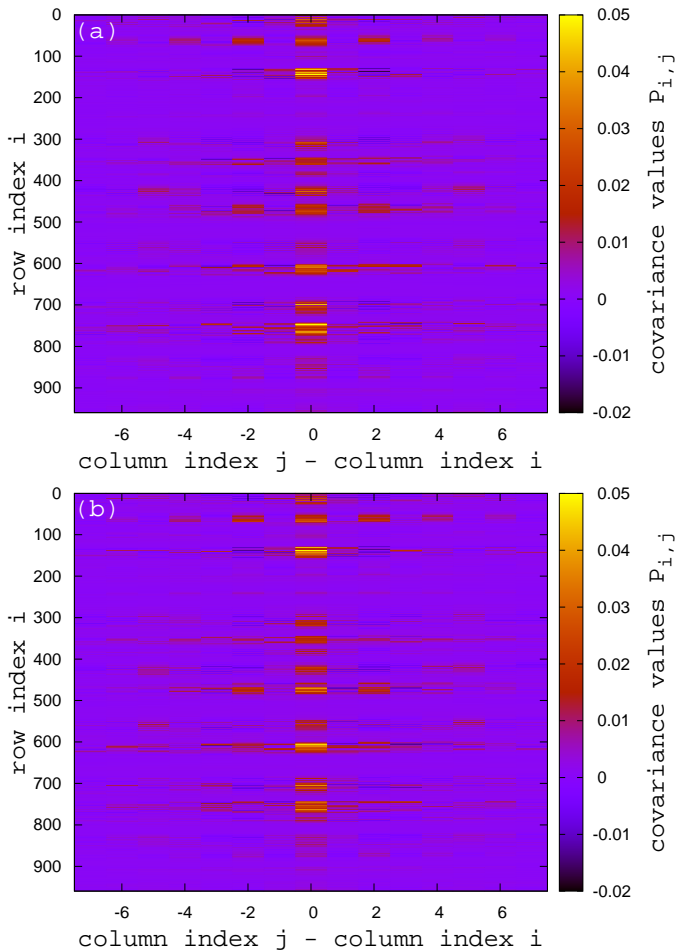}
\caption{Covariance matrix values of analysis ensemble at a certain forecast time of (a) ETKF and (b) LETKF. The $y$ axis is the row index $i$, and the $x$ axis is the difference between the column indices $j$ and $i$.}\label{fig-09}
\end{figure}

\begin{figure}
\noindent\includegraphics[width=21.18pc]{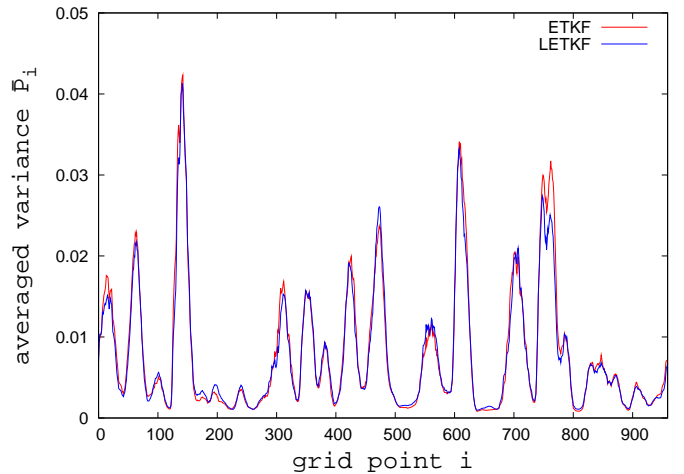}
\caption{Variance values of analysis ensemble at a certain forecast time. The values were averaged over 11 neighboring grid points. Red plot is for ETKF, and the blue plot is for LETKF.}\label{covariance-matrices-snapshot-6-analysis-diag}
\end{figure}

% covariance matrices
Now we compare the covariance matrices obtained from ETKF analyses with those obtained from LETKF analyses. We ran forecast cycles keeping two sets of ensembles, one for ETKF and one for LETKF. The number of ensemble members used in our ETKF implementation was 960, which is the total number of grid points of the model. Our LETKF implementation used a local patch size of $2 \times 50+1$ and number of ensemble members equal to 101, the same as the number of grid points in the local patch. The multiplicative covariance inflation factor was 1.13 for both ETKF and LETKF, which gives the minimum rms state estimate errors for both. We computed the covariance matrix of the ETKF analysis ensemble and the covariance matrix of the LETKF analysis ensemble at several analysis times. Because they have nonnegligible values only near the diagonal, we chose the elements $P_{i,j}$ of the covariance matrices with $|j-i|\le7$ and stacked them row by row so that elements with the same separation of indices $j-i$ are aligned vertically. Figures \ref{fig-09}a,b show the ETKF and LETKF covariance matrices at a representative analysis time. The vertical axis is index $i$, the horizontal axis is $j-i$, and the color represents the covariance values between grid points $i$ and $j$: $P_{i,j}$. Red corresponds to high values, and blue corresponds to low values. These figures show similar patterns for ETKF and LETKF. Figure \ref{covariance-matrices-snapshot-6-analysis-diag} is a plot of $\bar{P}_i$ versus $i$, where $\bar{P}_i$ is the average of $P_{i+l, i+l}$ over the 11 values of $l$ for $-5 \le l \le 5$. The red color is for ETKF, and the blue color is for LETKF. This figure also shows remarkably similar patterns for ETKF and LETKF. The mean rms errors of the state estimates [defined as $\langle \, \sqrt{\langle \, (\bar{Z}_n^a - Z_n)^2 \, \rangle_n} \, \rangle_{cycle}$, where $\bar{Z}_n^a$ is the mean of the analysis ensemble and $\langle \cdots \rangle_{n}$ and $\langle \cdots \rangle_{cycle}$ are averages over grid points and over forecast cycles, respectively] obtained from ETKF and LETKF are 0.0979 and 0.0997.

\begin{figure}
\noindent\includegraphics[width=21.18pc]{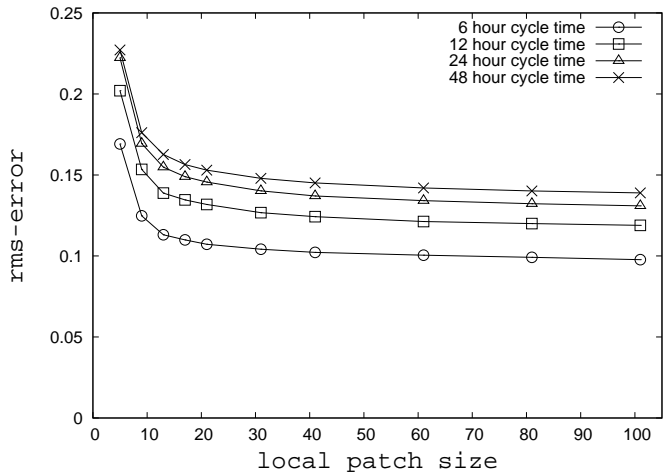}
\caption{Averaged rms errors of the state estimates by the means of the analysis ensembles vs local patch size. The values were averaged over 2000 forecast cycles. The lines with the circles, squares, triangles, and times signs correspond to 6-, 12-, 24-, and 48-h cycle times.}\label{var-diff-vs-patch-size-rms-error}
\end{figure}

\begin{figure}[!h]
\noindent\includegraphics[width=21.18pc]{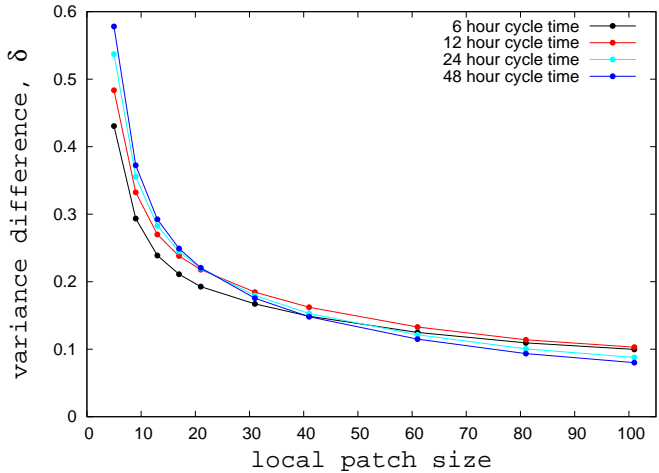}
\caption{Values of $\delta$ (normalized difference between variances of analysis ensembles of ETKF and LETKF) vs local patch size. The values were averaged over 2000 forecast cycles. The black, red, cyan, and blue colors correspond to 6-, 12-, 24-, and 48-h cycle times.}\label{var-diff-vs-patch-size-diff}
\end{figure}

\begin{figure}
\noindent\includegraphics[width=21.18pc]{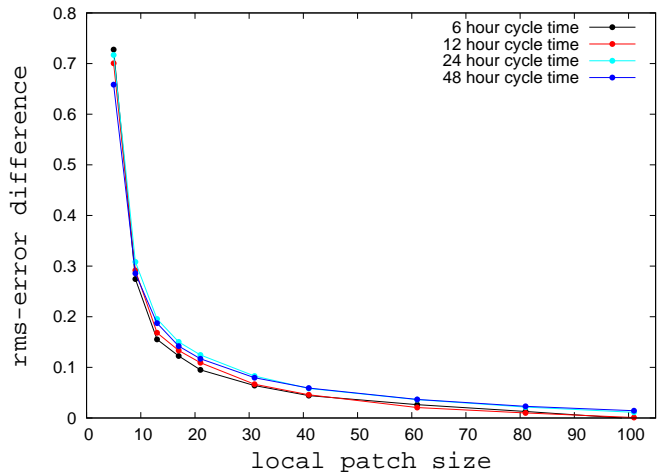}
\caption{Normalized difference between rms errors of state estimates with ETKF and LETKF vs local patch size. The values were averaged over 2000 forecast cycles. The black, red, cyan, and blue colors correspond to 6-, 12-, 24-, and 48-h cycle times.}\label{var-diff-vs-patch-size-rms-error-normalized}
\end{figure}

% measures of goodness of LETKF
We also studied the patch size and cycle time dependence of two global measures of the goodness of LETKF assimilations for 6-, 12-, 24-, and 48-h cycle times. One measure is the time-averaged rms error of the state estimate by averaged ensemble as defined above. The other measure is
\begin{equation}
\delta = \left< \, \frac{\left< | \bar{P}^{LETKF}_i - \bar{P}^{ETKF}_i | \right>_i}{\left< \bar{P}^{ETKF}_i \right>_i} \, \right>_{cycle},
\end{equation}
where $\langle \cdots \rangle_i$ denotes an average over all of the grid points. We used the same number of ensemble members as the number of grid points in a local patch for LETKF. Figures \ref{var-diff-vs-patch-size-rms-error} and \ref{var-diff-vs-patch-size-diff} show that the rms error and the quantity $\delta$ become smaller as we increase the local patch size. Using the rms error as our figure of merit, we see that there is relatively little benefit to increasing the patch size past about 15, although there is more substantial improvement in $\delta$ as the patch size is increased past 15.
Thus, we see that the improvement in the estimate of the mean of the ensemble saturates more quickly with increase in patch size than the improvement in the estimate of the variance of the ensemble.
From Fig.~\ref{var-diff-vs-patch-size-diff}, we can also see that longer forecast cycle time makes $\delta$ larger when the local patch size is smaller than about 20. However, we cannot see a clear pattern when the local patch size is large. Figure \ref{var-diff-vs-patch-size-rms-error-normalized} shows the difference between the rms errors of LETKF and ETKF divided by the rms error of ETKF, a normalized rms error difference.

\section{Demonstration of the localized influence of observations}\label{Demonstration of the localized influence of observations}

% localization does not lose much information
If the local patch size is large enough to encompass most of the correlation seen in Fig.~\ref{correlations-ensemble-06h}, then it is to be expected that the analysis at grid points around the middle of the patch does not lose too much information by ignoring observations outside the local patch. In what follows, we report evidence supporting this intuition.

\begin{figure}[!h]
\noindent\includegraphics[width=21.18pc]{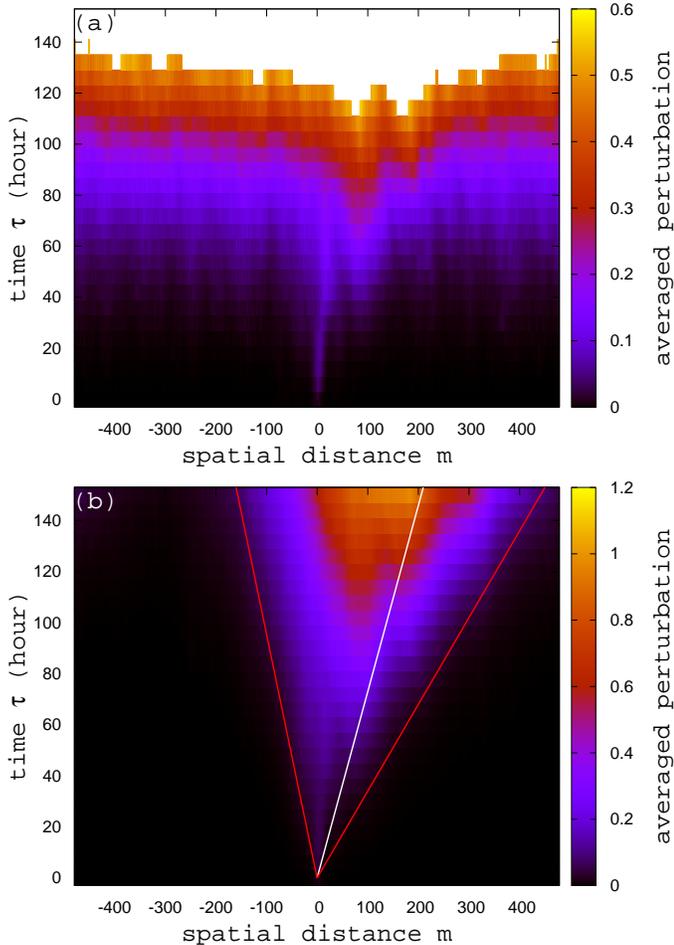}
\caption{Values of $\langle D_m(\tau) \rangle_{cycle}$ (evolution of the perturbation to the ensemble mean caused by five localized observations). The values were averaged over 4000 cycles, (a) with ETKF and (b) with LETKF. The white line shows the group velocity of Lorenz model 3, and the two red lines show the boundaries of the wedgelike region of the perturbations.}\label{fig-14}
\end{figure}

% Effect of local observations on other points in different space-time
We now examine how observations taken at time $t$ and in a small region around spatial location $i$ affect a forecast at space-time point $(t+\tau,i+m)$. To do this, we first ran 100 forecast cycles using the ETKF and noisy observations at all the grid points (as done in section \ref{Spatial correlations of the background ensemble}), thus bringing the ensemble members close to the true state. Then, at current time $t$, we took observations at only five fixed contiguous locations, $i-2,i-1,i,i+1$, and $i+2$ (taking observations at only one location did not produce a clear result, possibly because the impact of one observation was not strong enough). We then used these observations to calculate an analysis ensemble. Then, we calculated the mean of the background ensemble and the mean of the analysis ensemble and evolved the two means from the current time $t$ to the time $t+\tau$. We denote the values of these two evolved means at each grid point $n$ by $\bar{Z}_n^b(\tau)$ and $\bar{Z}_n^a(\tau)$ ($b$ and $a$ denote background and analysis, respectively). We did this for values of $\tau$ ranging from $\tau=0$ h to $\tau=150$ h in 6-h intervals, calculating and saving the quantities
\begin{equation}
D_m(\tau)=( \, \bar{Z}_{i+m}^a(\tau) - \bar{Z}_{i+m}^b(\tau) \, )^2.
\end{equation}
We then proceed to the next time cycle $t \rightarrow t + 6$ h and repeat this process: i.e., at time $t$ we now take observations at all the grid points, calculate the analysis ensemble, evolve it and the true state forward to create a new background ensemble at the new time $t \rightarrow t + 6$ h, and repeat the previously described five-observation-point calculation of $D_m(\tau)$. Figure \ref{fig-14}a shows a plot of $\langle D_m(\tau) \rangle_{cycle}$, the average of $D_m(\tau)$ over 4000 cycles. The horizontal axis is $m$, and the vertical axis is $\tau$. The color represents $\langle D_m(\tau) \rangle_{cycle}$. Red corresponds to high values, and blue corresponds to low values. The white color represents values higher than the upper limit of the color bar. The figure shows that observations do not affect points that are far from the observation points within up to about 30 h. However, all of the points become affected by the observations after about 30 h.

\begin{figure}[!h]
\noindent\includegraphics[width=21.18pc]{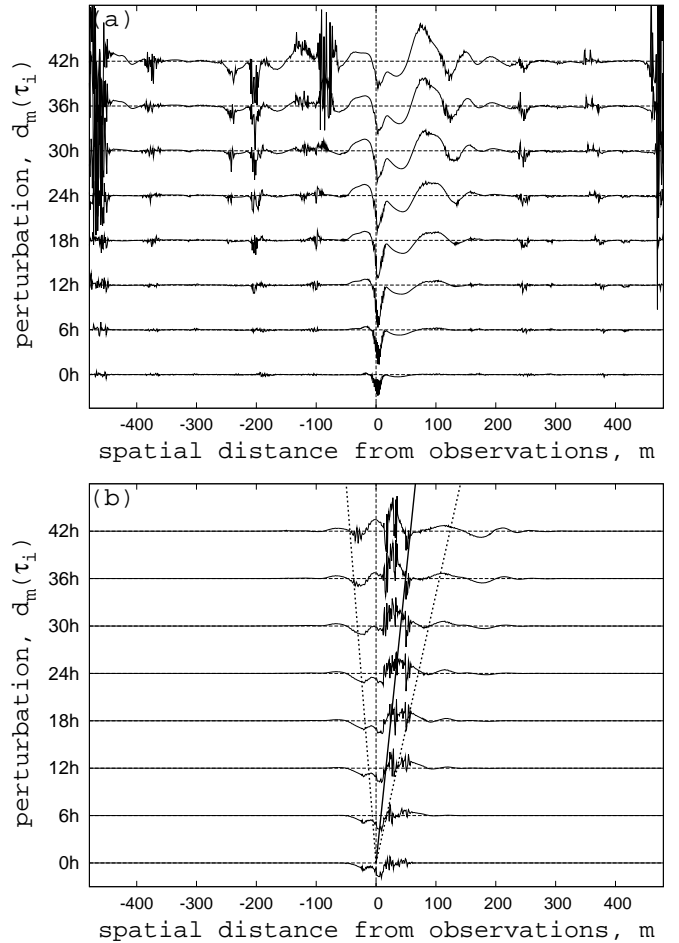}
\caption{Evolution of the perturbation to the ensemble mean caused by five localized observations, $d_{m}(\tau_i)$, in 6-h time intervals from $\tau=0$ h to $\tau=42$ h, at a certain cycle time, (a) with ETKF and (b) with LETKF. The solid line shows the group velocity of Lorenz model 3, and the two dotted lines show the same boundaries in Fig.~\ref{fig-14}b.}\label{fig-15}
\end{figure}

\begin{figure}[!h]
\noindent\includegraphics[width=21.18pc]{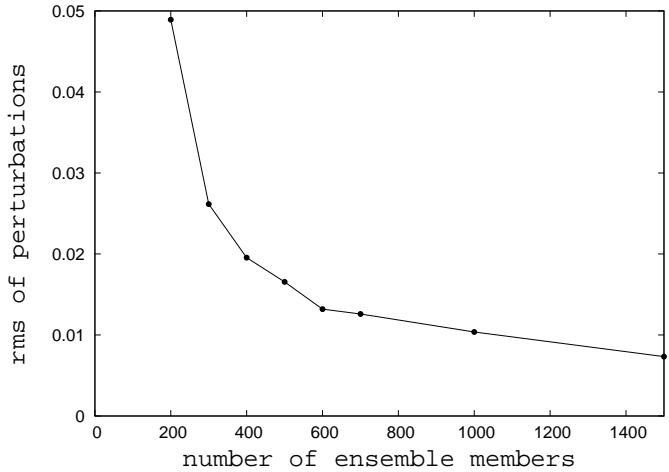}
\caption{The rms values of $d_m(0)$ divided by the size of the noise of observations vs ensemble size of ETKF. The values were averaged over 2000 cycles and over grid points that are more than 300 grid points away from the middle of the observation points.}\label{noisy-impact}
\end{figure}

% evolution of difference for one instance
To investigate why all of the points eventually become affected after some time, we plotted the time evolution of the differences between the two means,
\begin{equation}
d_{m}(\tau_i) = \bar{Z}_{i+m}^a(\tau_i) - \bar{Z}_{i+m}^b(\tau_i),
\end{equation}
at $\tau_i = i \times 6$ h ($i=0,1, \ldots, 7$) at a certain cycle time (see Fig.~\ref{fig-15}a). Figure \ref{fig-15}a shows $d_{m}(\tau_i)$ plotted versus $m$ at 6-h intervals from $\tau=0$ h to $\tau=42$ h.
It is seen that small differences at points far from the observation points at $\tau=0$ h become amplified and dominate the $d_m$ values at later times. At small $\tau$, these differences are close to zero but are not exactly zero. We computed the rms values of $d_{m}(0)$ averaged over grid points $m$ and cycles varying the number of ensemble members, where the average over $m$ is taken only for $m<-300$ and $m>300$. Figure \ref{noisy-impact} shows that the impact of the localized observation on points far from the observation point decreases as the ensemble size increases.
So, these small impacts are likely to be caused by the finite size of the ensemble [see \cite{Hamill-et-al-01} and \cite{Anderson-07} for detailed discussions of noisy covariance values]. As $\tau$ increases, the small perturbations far from the observation points are chaotically amplified and become large. We repeated this experiment using LETKF in place of ETKF. The use of localization in LETKF has the effect of eliminating the randomlike small initial differences $d_m(0)$ at large $|m|$. Figures \ref{fig-14}b and \ref{fig-15}b show the results. It is seen that the effect of the localized observations spreads to other points linearly, creating a wedgelike region in our $\tau$-versus-$m$ diagram indicated by two red lines in Fig.~\ref{fig-14}b and two dotted lines in Fig.~\ref{fig-15}b. The propagation speed corresponding to the right boundary of the wedge is roughly 3 grid points per hour, and the propagation speed corresponding to the left boundary of the wedge is roughly 1 grid point per hour. Thus, it appears that, during a 6-h cycle time, the effect of an observation will, on average, reach up to about 18 grid points to the right and up to about 6 grid points to the left. The maximum difference between the background mean and the analysis mean propagates at a speed that is close to the group velocity of wave packets (1.37 grid points per hour, shown by the white line in Fig.~\ref{fig-14}b and the solid line in Fig.~\ref{fig-15}b) determined from Fig.~\ref{envelope}.

\begin{figure}[!h]
\noindent\includegraphics[width=21.18pc]{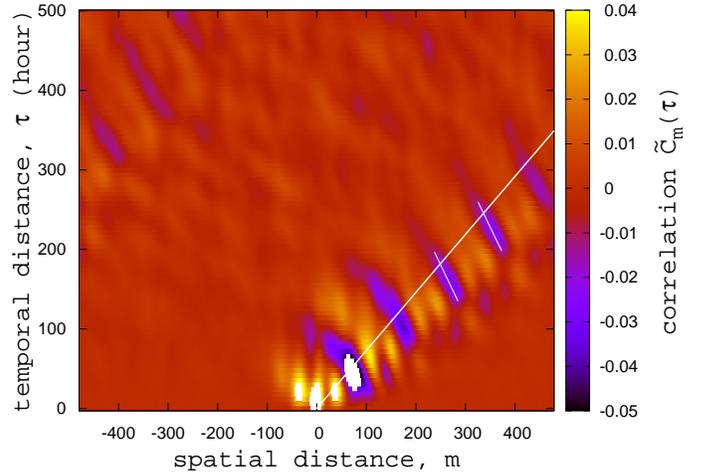}
\caption{Values of $\tilde{C}_m(\tau)$ [correlations between perturbations at $(0, n)$ and at $(\tau, n+m)$ caused by observations]. The observations were taken at all the grid points. The values were averaged over grid points $n$ and forecast cycles. The long white line shows the group velocity of the Lorenz model 3, and the short white lines show phase velocity of the model.}\label{effect}
\end{figure}

% Effect of global observations on other points in different space-time
To see the effect of observations at a space-time point $(t,n)$ on a forecast at $(t+\tau,n+m)$ when observations are taken at all of the grid points, we did an experiment similar to the one described above. This time, we took observations at all of the grid points (instead of at 5 adjoining grid points) and calculated the value $\tilde{C}_m(\tau)$ instead of $D_m(\tau)$ defined in the above experiment:
\begin{equation}
\tilde{C}_m(\tau) = \left< \frac{[\bar{Z}_{n+m}^a(\tau) - \bar{Z}_{n+m}^b(\tau)] [\bar{Z}_{n}^a(0) - \bar{Z}_{n}^b(0)]}{\sigma_{noise}^2} \right>_{n,cycle},
\end{equation}
where $\sigma_{noise}$ denotes the standard deviation of the observation noise. Figure \ref{effect} shows $\tilde{C}_m(\tau)$.
The horizontal axis is $m$, and the vertical axis is $\tau$. The color represents $\tilde{C}_m(\tau)$. Red corresponds to high values, and blue corresponds to low values. The white color represents values that are higher than the upper limit of the color bar. This plot also shows that the effect of observations propagates globally to the right at a speed that is close to the group velocity of wave packets (shown by a long white line). The orientation of major axes of the blue blobs slopes to the left, and this slope is on the order of the phase velocity (-0.77 grid points per hour, shown by short white lines) determined from Fig.~\ref{correlations}.

\section{Conclusions}\label{Conclusion}

% Conclusions
The correlation function obtained from the time evolution of the state of Lorenz model 3 and the envelopes extracted from it both show that there is predominant wave packet propagating to the right (sections \ref{Model} and \ref{Correlation structure of the model}). We found that the correlation length of the deviations from the ensemble mean of the background ensemble of the ETKF at each forecast cycle with Lorenz model 3 is very much shorter than the correlation length of the climatological distribution of model states (section \ref{Spatial correlations of the background ensemble}). Thus, we argued that we do not lose much information from localization of the analysis if the size of the local patch is big enough to cover the correlation length of the deviations of the background ensemble. The comparison of the covariance matrices of the analysis ensembles with and without localization shows that they have similar patterns (section \ref{ETKF and LETKF covariance matrices}), thus providing strong support for the achievable accuracy of localization. The effect of an observation at space-time point $(t,n)$ on a forecast at $(t+\tau,n+m)$ was found to be local. In addition, the information obtained from the observations propagates both forward (to the right) and backward (Fig.~\ref{fig-14}b). However, the forward propagation, which is in the direction of the group velocity (to the right), is faster than the backward propagation (to the left), and the maximum effect propagates at a speed that is close to the group velocity (section \ref{Demonstration of the localized influence of observations}).

\begin{acknowledgment} 
This work was supported by ONR contract N00014-07-1-0734 and NSF contract ATM-0935538. Craig Bishop and an anonymous reviewer made several useful suggestions to improve the paper.
\end{acknowledgment}

% Use appendix}[A], {appendix}[B], etc. etc. in place of appendix if you have multiple appendixes.
\ifthenelse{\boolean{dc}}
{}
{\clearpage}
\begin{appendix}[A]
\section*{\begin{center}Extracting an envelope\end{center}}

\begin{figure}
\noindent\includegraphics[width=21.18pc]{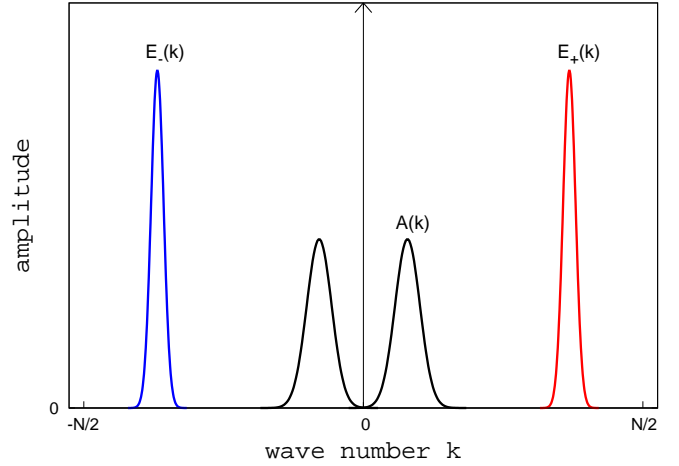}
\caption{DFT of $a(n)$, $e^{i\phi(n)}$, and $e^{-i\phi(n)}$. The terms $A(k)$, $E_{+}(k)$, and $E_{-}(k)$ are DFTs of $a(n)$, $e^{i\phi(n)}$, and $e^{-i\phi(n)}$, respectively, and $f(n) = a(n) \cos[\,\phi(n)\,], \, 0 \le n < N$.}\label{diagram-k}
\end{figure}

See \cite{Zimin-et-al-03} for the original introduction of the method presented here. Let us say that we have a function
\begin{equation}
f(n) = a(n) \cos[\,\phi(n)\,], \, 0 \le n < N,
\end{equation}
where $n$ is an integer, $a(n)$ is an envelope, and $\cos[\,\phi(n)\,]$ is an oscillating part with $\phi(n)$ monotonically increasing. We want to extract $a(n)$ given $f(n)$. First, $f(n)$ can be expressed as
\begin{equation}
f(n) = \frac{1}{2} a(n) e^{i\phi(n)} + \frac{1}{2} a(n) e^{-i\phi(n)}.
\end{equation}
Define $F(k)$, $A(k)$, $E_{+}(k)$, and $E_{-}(k)$ to be discrete Fourier transforms (DFTs) of $f(n)$, $a(n)$, $e^{i\phi(n)}$, and $e^{-i\phi(n)}$, respectively. Assuming each function is N-periodically extended, we have the following:
\begin{align}
F(k) &= \frac{1}{2} \, \text{DFT}\{ a(n) e^{i\phi(n)} \} + \frac{1}{2} \, \text{DFT}\{ a(n) e^{-i\phi(n)} \} \\
	 &= \frac{1}{2N} A(k) * E_{+}(k) + \frac{1}{2N} A(k) * E_{-}(k) \\
	 &\equiv \frac{1}{2N} \sum_{l=0}^{N-1} A(l)E_{+}(k-l) + \frac{1}{2N} \sum_{l=0}^{N-1} A(l)E_{-}(k-l),
\end{align}
where the asterisk denotes circular convolution. The term $E_{+}(k)$ is centered around a positive frequency, and $E_{-}(k)$ is centered around a negative frequency. If we assume that $a(n)$ changes much more slowly than $\phi(n)$, then $A(k)$ is well separated from $E_{+}(k)$ and $E_{-}(k)$. See Fig.~\ref{diagram-k}, where the horizontal axis is the wavenumber $k$ and the vertical axis is the amplitude of the DFT values. The red, blue, and black plots are for DFTs of $e^{i\phi(n)}$, $e^{-i\phi(n)}$, and $a(n)$, respectively. Therefore, $A(k) * E_{+}(k)$ will be on the positive frequency side and $A(k) * E_{-}(k)$ will be on the negative frequency side. So, taking inverse discrete Fourier transform of $F(k)$ only with positive frequencies, $\text{DFT}^{-1}_{+}\{ F(k) \}$, is equivalent to taking inverse discrete Fourier transform of $(2N)^{-1} A(k) * E_{+}(k)$. Therefore, we have the following:
\begin{align}
\text{DFT}^{-1}_{+}\{ F(k) \} &= \text{DFT}^{-1}\left\{ \frac{1}{2N} A(k) * E_{+}(k) \right\} \\
&= \frac{1}{2} a(n) e^{i\phi(n)}
\end{align}
Thus,
\begin{equation}
a(n) e^{i\phi(n)} = 2 \, \text{DFT}^{-1}_{+}\{ F(k) \}
\end{equation}
Taking the absolute values of both sides, we obtain
\begin{equation}
a(n) = 2 \, | \, \text{DFT}^{-1}_{+} \, \{ F(k) \} |.
\end{equation}
In the above equation, only several frequencies that have large DFT values can be used to approximate the envelope. This has the effect of smoothing out the envelope by sacrificing accuracy. As more and more frequency components are added, more and more wiggles are present in the envelope.
\end{appendix}

\ifthenelse{\boolean{dc}}
{}
{\clearpage}
\begin{appendix}[B]
\section*{\begin{center}LETKF algorithm\end{center}}
We present a summary of the LETKF algorithm of \cite{Hunt-et-al-07} used in our study with Lorenz model 3. We assume a discrete model that has grid points. Each grid point has a variable. A local region associated with grid point $n$ is defined as a region consisting of grid points $i$ with $n-L \le i \le n+L$ for a certain integer $L$.\\\\
\textbf{Input}: a global background ensemble of $m_g$-dimensional model state vectors $\{\mathbf{x}^{b(i)}_g,i=1,2,\ldots,k\}$, an $l_g$-dimensional vector $\mathbf{y}^o_g$ of observations, a function $\mathbf{H}$ that maps the $m_g$-dimensional model space to the $l_g$-dimensional observation space, and an $l_g \times l_g$ observation error covariance matrix $\bm{\mathsf{R}}_g$. Here the subscript $g$ denotes global.\\
\textbf{Output}: a global analysis ensemble of $m_g$-dimensional model state vectors $\{\mathbf{x}^{a(i)}_g,i=1,2,\ldots,k\}$\\
\textbf{1}. Apply $\mathbf{H}$ to each $\mathbf{x}^{b(i)}_g$ to form the global background observation ensemble $\{\mathbf{y}^{b(i)}_g\}$ and average the latter vectors to get the $l_g$-dimensional column vector $\bar{\mathbf{y}}^b_g$. Subtract this vector from each $\mathbf{y}^{b(i)}_g$ to form the columns of the $l_g \times k$ matrix $\bm{\mathsf{Y}}^b_g$.\\
\textbf{2}. Average the vectors $\{\mathbf{x}^{b(i)}_g\}$ to get the $m_g$-dimensional column vector $\bar{\mathbf{x}}^b_g$ and subtract this vector from each $\mathbf{x}^{b(i)}_g$ to form the columns of the $m_g \times k$ matrix $\bm{\mathsf{X}}^b_g$.\\
\\
For each grid point, do the following steps \textbf{3}--\textbf{8}.\\
\\
\textbf{3}. Select the rows of $\bar{\mathbf{x}}^b_g$ and $\bm{\mathsf{X}}^b_g$ corresponding to the local region of $m$ grid points associated with the given grid point, forming the $m$-dimensional vector $\bar{\mathbf{x}}^b$ and the $m \times k$ matrix $\bm{\mathsf{X}}^b$. Select the rows of $\bar{\mathbf{y}}^b_g$ and $\bm{\mathsf{Y}}^b_g$ corresponding to the $l$ observations for the local region, forming the $l$-dimensional vector $\bar{\mathbf{y}}^b$ and the $l \times k$ matrix $\bm{\mathsf{Y}}^b$. Select the corresponding rows of $\mathbf{y}^o_g$, forming the $l$-dimensional vector $\mathbf{y}^o$. Select the corresponding rows and columns of $\bm{\mathsf{R}}_g$, forming the $l \times l$ matrix $\bm{\mathsf{R}}$.\\
\textbf{4}. Compute the $k \times l$ matrix $\bm{\mathsf{C}}=(\bm{\mathsf{Y}}^b)^T \bm{\mathsf{R}}^{-1}$.\\
\textbf{5}. Compute the $k \times k$ matrix $\tilde{\bm{\mathsf{P}}}^a = [(k-1)\bm{\mathsf{I}}/\rho+\bm{\mathsf{C}}\bm{\mathsf{Y}}^b]^{-1}$, where $\rho > 1$ is a covariance inflation factor.\\
\textbf{6}. Compute the $k \times k$ matrix $\bm{\mathsf{W}}^a = [(k-1)\tilde{\bm{\mathsf{P}}}^a]^{1/2}$, where the power $1/2$ means the symmetric square root.\\
\textbf{7}. Compute the $k$-dimensional vector $\bar{\mathbf{w}}^a = \tilde{\bm{\mathsf{P}}}^a\bm{\mathsf{C}}(\mathbf{y}^o-\bar{\mathbf{y}}^b)$ and add it to each column of $\bm{\mathsf{W}}^a$, forming a $k \times k$ matrix whose columns are the analysis vectors $\{\mathbf{w}^{a(i)}\}$.\\
\textbf{8}. Multiply $\bm{\mathsf{X}}^b$ by each $\mathbf{w}^{a(i)}$ and add $\bar{\mathbf{x}}^b$ to get the local analysis ensemble members $\{\mathbf{x}^{a(i)}\}$ for the given grid point.\\
\\
\textbf{9}. After performing steps 3--8 for each grid point, form the global analysis ensemble $\{\mathbf{x}^{a(i)}_g\}$ by combining $\{\mathbf{x}^{a(i)}\}$'s from each local region by taking weighted averages, $x^{a(i)}_{n,g} = \sum_{j=n-L}^{n+L} x^{a(i)}_{n,l(j)} f(n-j)$, where the subscript $l(j)$ denotes the local region associated with the grid point j and $f(\cdot)$ is a weighting function defined on the integers from $-L$ to $L$ for a certain value $L$.
\end{appendix}

% Create a bibliography directory and place your .bib file there.
\ifthenelse{\boolean{dc}}
{}
{\clearpage}
%\bibliographystyle{./ametsoc}
%\bibliography{../references/references}

\end{document}